\documentclass[aps,prb,twocolumn,superscriptaddress,showpacs]{revtex4}

\usepackage[dvips]{graphicx}

\begin{document}

\title{Dimensional Tuning of the Magnetic-Structural Transition in A(Fe$_{1-x}$Co$_x$)$_2$As$_2$ (A=Sr,Ba)}

\author{Jack Gillett}
\affiliation{Cavendish Laboratory, J. J. Thomson Ave, University of Cambridge, UK}

\author{Sitikantha D. Das}
\affiliation{Cavendish Laboratory, J. J. Thomson Ave, University of Cambridge, UK}

\author{Paul Syers}
\affiliation{Cavendish Laboratory, J. J. Thomson Ave, University of Cambridge, UK}

\author{Alison K. T. Ming}
\affiliation{Cavendish Laboratory, J. J. Thomson Ave, University of Cambridge, UK}

\author{Jose I. Espeso}
\affiliation{Dpto. CITIMAC, Universidad de Cantabria, Spain}

\author{Chiara M. Petrone}
\affiliation{Department of Earth Sciences, University of Cambridge, UK}

\author{Suchitra E. Sebastian}
\affiliation{Cavendish Laboratory, J. J. Thomson Ave, University of Cambridge, UK}

\date{\today}

\begin{abstract}

A phase diagram of superconducting Sr(Fe$_{1-x}$Co$_x$)$_2$As$_2$ as a function of doping ($x$) is determined by a series of thermodynamic and transport measurements on single crystals. On comparison with a similar phase diagram for Ba(Fe$_{1-x}$Co$_x$)$_2$As$_2$ (Co-doped Ba122), we find that the increased dimensionality of Co-doped Sr122  results in a single first-order-like transition where the magnetic and structural transitions coincide, unlike the case of Co-doped Ba122 that exhibits split quasicontinuous magnetic and structural transitions. We relate this dimensionally-tuned splitting in the magnetic and structural transitions to the relative size of superconducting temperatures in these materials.

\end{abstract}


\maketitle

The discovery of the new family of high-\it T\rm $\rm _c$ Iron Pnictide superconductors provides an opportunity to better understand factors that enhance superconductivity. A striking feature in common between Iron Pnictide, Cuprate, and Heavy Fermion superconductors is the proximity of superconductivity to magnetism. In the $R$FeAsO (known as 1111, where $R$ is a rare earth element)~\cite{laofeas} and $A$Fe2As2 (known as 122, where $A$ is an alkaline earth element)~\cite{kdopedba} families of Iron Pnictide superconductors, the undoped material is an antiferromagnet which evolves into a superconductor upon doping. The manner in which magnetism evolves into superconductivity is of particular interest, especially whether a quasicontinuous magnetic phase transition is involved. The 122 system has attracted interest due to the appearance of superconductivity either by chemical (hole-~\cite{kdopedba} or electron-~\cite{codopedba2}$^,$~\cite{nickel}$^,$~\cite{copper}$^,$~\cite{rhodium}) doping; or by applied pressure~\cite{pressuresr3}; and the relative ease with which large crystals can be grown. Co-doped Ba122 has been well studied~\cite{pressureba}$^,$~\cite{phasediagramba}$^,$~\cite{pressureba}$^,$~\cite{phasediagramba2}; exhibiting superconducting temperatures up to $\approx$24K. The single magnetic transition in the undoped compound is observed to split into magnetic and structural transitions at higher doping and followed into the superconducting state via x-ray experiments, with superconductivity found in both the orthorhombic and tetragonal states~\cite{reentrytetragonal}.

In this paper we study the interplay of magnetism and superconductivity in Co-doped Sr122 [ref.~\onlinecite{codopedsr}], and compare this with the Ba122 family. We perform a careful examination of the evolution of the magnetic to superconducting transition with doping, and in particular explore whether any splitting of the structural and magnetic transitions can be discerned in the manner of Co-doped Ba122. We find evidence of a largely first order transition in Sr122 where the magnetic transition coincides with the structural transition. The first orderness of this transition is related to the larger interlayer exchange in the more three-dimensional~\cite{firstsr122} Sr122 and associated with the suppressed superconductivity, and further separation of the superconducting dome in Co-doped Sr122 compared to Co-doped Ba122. The contribution of magnetic interactions at a quasicontinuous transition to the enhancement of superconductivity in the family of 122 superconducting pnictides is suggested.

Single crystals of Co-doped Sr122 and Co-doped Ba122 of up to 5mm $\times$ 5mm $\times$ 0.5 mm were grown using a flux technique described in previous work~\cite{qosr}. FeAs was synthesized by sealing Fe (powder, 99.998\%, Alfa Aesar) and As (lumps, 99.9999\%, Alfa Aesar) and heating to 1000 $^{\circ}$C gradually with intermediate dwells at 500 $^{\circ}$C and 800 $^{\circ}$C before cooling. The precursor material formed was then ground into powder and the procedure repeated to yield small silver-coloured lumps of highly stoichiometric FeAs. CoAs (99.5\%, Cerac), FeAs (pre-reacted) and Sr/Ba (pieces, 99.9\%, ESPI) were mixed in an argon filled glove box according to the ratio Sr:FeAs:CoAs = 1:4(1-$x$):4$x$, placed in an alumina crucible; covered by a catch crucible containing quartz wool and sealed in a quartz ampoule under 1/3 bar of argon. The ampoule was gradually heated to 1180 $^{\circ}$C followed by slow cooling to 1020 $^{\circ}$C at 3-4 $^{\circ}$C/hr, at which temperature the ampoule was centrifuged to remove the liquid flux, crystals retrieved and stored under vacuum. Crystals with nominal doping levels of Co in the range 0.00 $<$ $x$ $<$ 0.19 were grown.

\begin{figure}
\includegraphics[width=3 in]{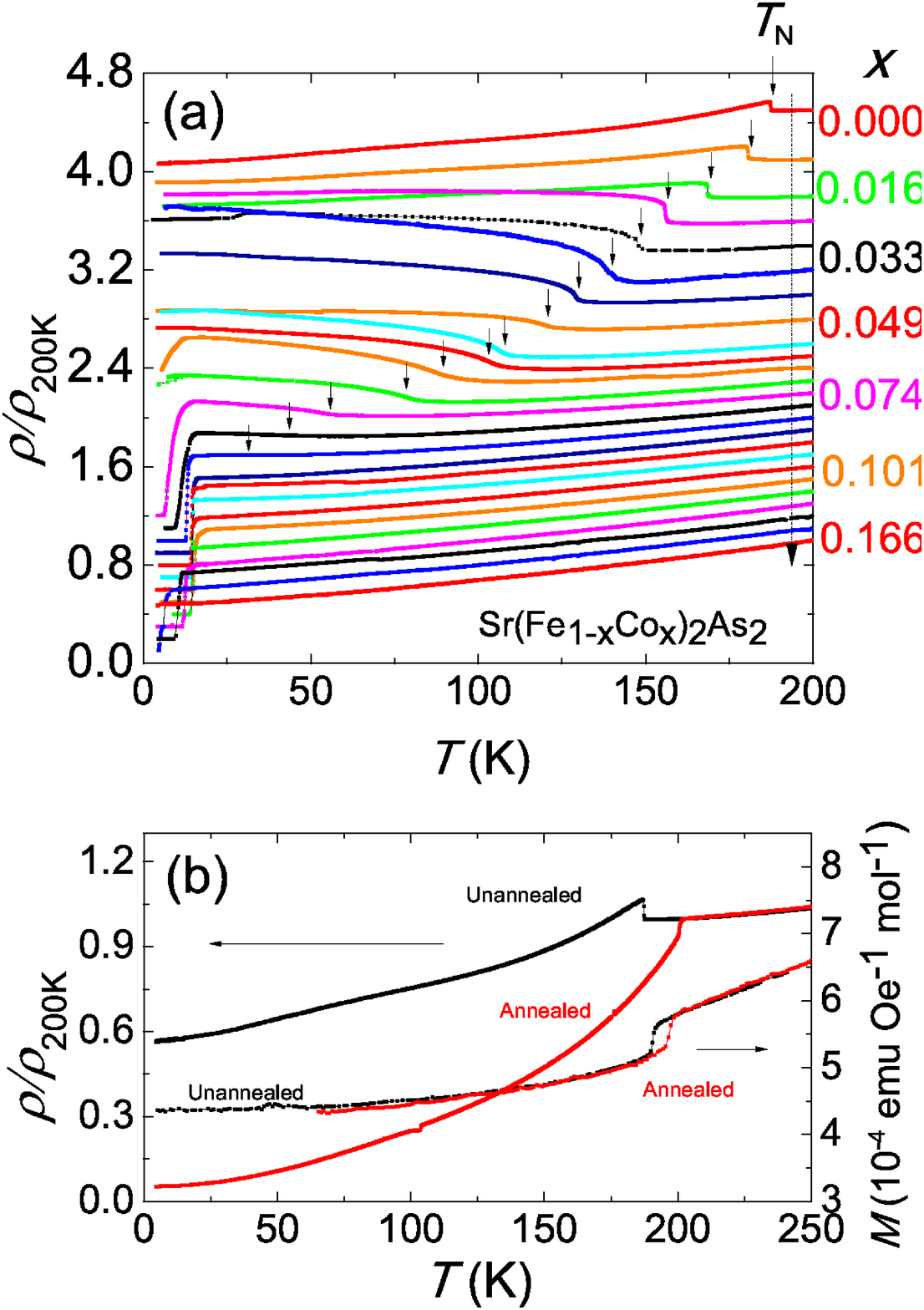}
 \caption{\label{fig:resistivity} (colour online) The electrical resistivity ($\rho$/$\rho_{\rm 200K}$) as a function of doping in Co-doped Sr122 from $x=0$ to $x=0.166$. Data is normalised to the resistivity above the Neel transition in all samples, and for clarity, data for the various dopings is offset. $T_{\rm N}$ (defined as the point of greatest negative slope in the resistance) is indicated by arrows. b) shows the effect of annealing on a typical undoped Sr122 crystal. $R_{200\rm K}$/$R_{4\rm K}$ is improved and $T_{\rm N}$ is increased. Magnetisation also shows the increase in $T_{\rm N}$. Magnetisation data for annealed sample is offset by a constant for comparison with unannealed data.}
\end{figure}

Heat capacity measurements were made using a relaxation method on a commercial Quantum Design (QD) PPMS, DC magnetisation was carried out on a QD MPMS, and in-plane resistance measurements were carried out on a custom-built helium flow cryostat using an excitation current of 50$\mu$A at a frequency of 77.7Hz using the standard four-point technique. Sample doping was determined by electron probe microanalysis on a Cameca SX100 equipped with 1 EDS and 5 WDS spectrometers based in the Department of Earth Sciences by measuring 20 points per sample. Variation within a sample was found to be small and almost independent of doping, remaining less than 2\% for the majority of samples, and $\sim$10\% for the lowest doping. No impurity phases were observed. Annealing was performed by heating to 850 $^{\circ}$C followed by a 48 hr dwell under argon or in vacuum, followed by cooling to room temperature.

\begin{figure}
\includegraphics[width=3 in]{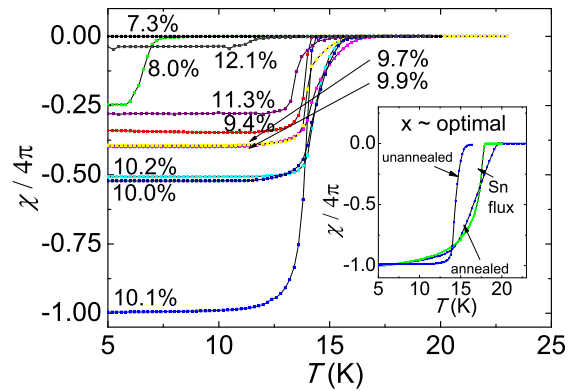}
 \caption{\label{fig:magnetisation} (colour online) Shielding in DC magnetisation corresponding to the development of superconductivity with doping ($\chi$ = $M$/$H$). All measurements were performed in a 50 Oe field applied parallel to the c-axis of the crystal after ZFC. The volume fraction reflects bulk SC in all cases, and has been renormalised to 1 for the highest volume fraction for purposes of comparison. $T_{\rm c}$ is defined as the point of greatest negative slope in the magnetisation. Inset: magnetisation for annealed and Sn-grown cobalt-doped samples with close to optimal $x$ show higher onset \it T\rm $\rm _c$ and a broader transition compared to the unannealed sample (volume fraction has been renormalised to be equal).}
\end{figure}

\begin{figure*}
\includegraphics[width=7 in]{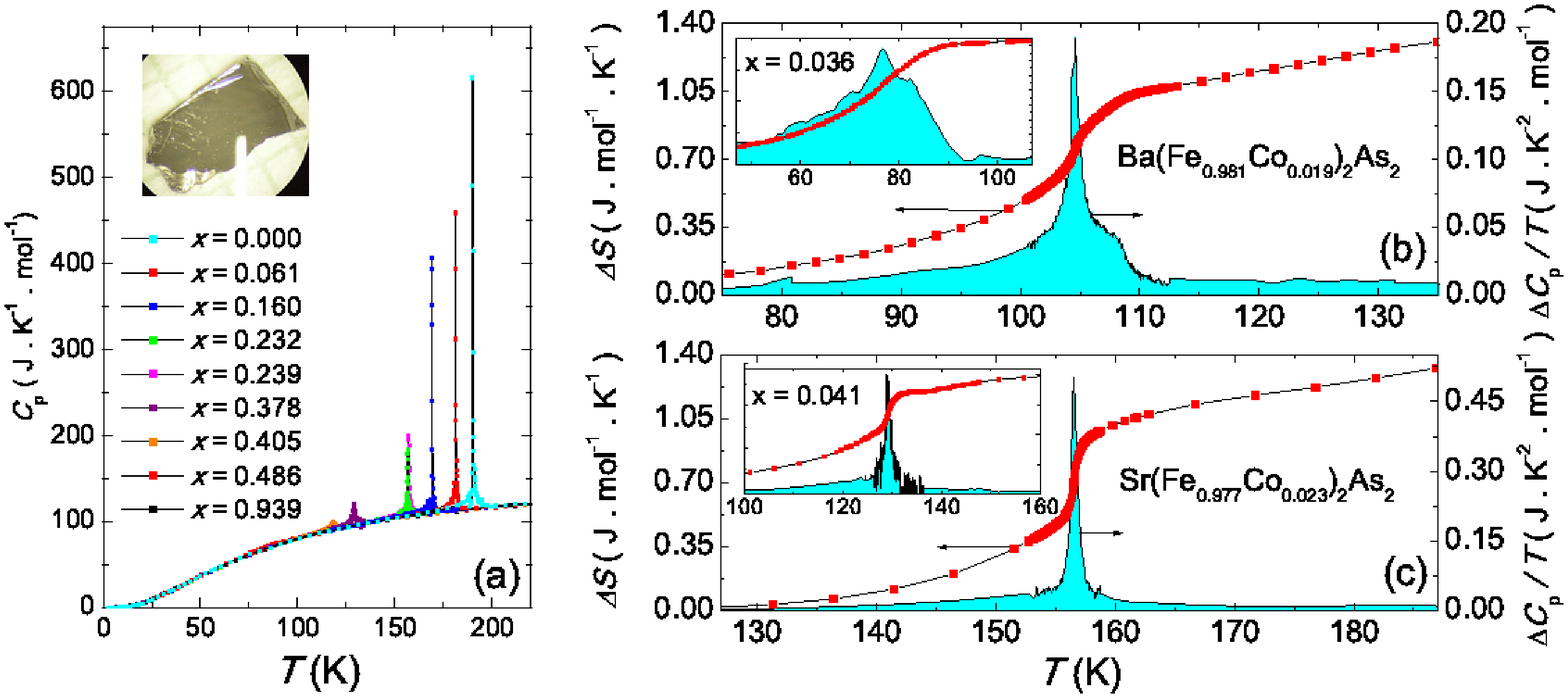}
 \caption{\label{fig:heatcapacity} (colour online) a) shows the heat capacity for a series of Co-doped Sr122 samples. Because of the narrowness of the transition, a temperature rise of only 0.3\% was used for each measurement in the vicinity of the transition. The magnitude of the heat capacity jump at the transition falls with doping until it is unobservable at $x$=0.063. The inset shows a typical cut and cleaved crystal used in our measurements on a mm scale. b) and c) show a comparison of the entropy and heat capacity jumps for 1.9\% Co-doped Ba122 and 2.3\% Co-doped Sr122, as described in the text. Dopings for comparison of Ba122 and Sr122 are chosen to correspond to a similar percentage suppression of \it T\rm $\rm _N$. The transition in Co-doped Sr122 is seen to be much sharper and first-order-like than in Co-doped Ba122. The insets are similar data for higher doped 3.6\% Co-doped Ba122~(data taken from ref.~\onlinecite{phasediagramba}) and 4.1\% Co-doped Sr122 respectively, showing increased broadening in the Ba122, while the Sr122 remains sharp and first-order-like.}
\end{figure*}

Undoped samples of Sr122 show a structural and antiferromagnetic transition at \it T\rm $\rm _N$=192K (figure \ref{fig:resistivity}). On Co-doping, this transition is suppressed, with \it T\rm $\rm _N$ eventually disappearing, accompanied by the onset of superconductivity at $x$=0.075, extending from $x$ = 0.075 to 0.144, with the optimal superconducting temperature occurring at \it T\rm $\rm _c\approx$~16K for $x$=0.102.

\begin{figure}
\includegraphics[width=3 in]{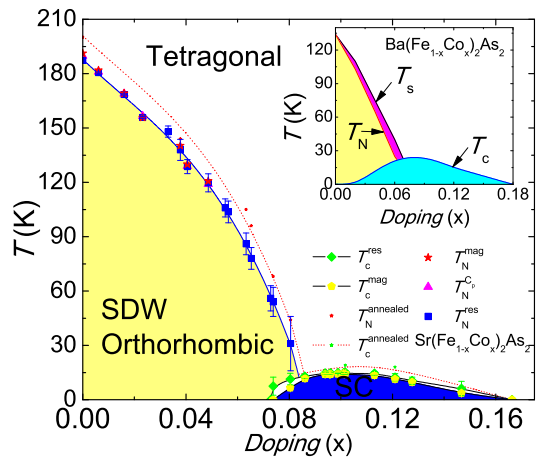}
 \caption{\label{fig:phasediagram} (colour online) The phase diagram of Co-doped Sr122 measured from Magnetisation, Heat Capacity and Resistivity. The upper dashed lines correspond to higher Neel and critical temperatures for annealed samples. The inset shows a schematic of the phase diagram of Co-doped Ba122 constructed from our measurements and previous reports~\cite{phasediagramba}$^,$~\cite{phasediagramba2}. The purple region between the structural and magnetic transitions that occurs in Co-doped Ba122 is not seen in Co-doped Sr122}
\end{figure}

Annealing is seen to have a rather interesting effect on Sr122 samples. The chief effect of annealing by the procedure described above is to improve the in-plane sample resistivity. The residual resistance ratio (RRR), defined as $R_{200\rm K }$/$R_{4\rm K}$, increases with annealing (for example, an increase from $\approx$~2 to $\approx$~20 is seen in figure \ref{fig:resistivity}b). Annealing also increases \it T\rm $\rm _N$ from 192K to 200K in undoped samples along with the suppression in the upturn in the resistance at the antiferromagnetic transition before a rapid drop in resistance at lower temperatures. In doped samples, \it T\rm $\rm _N$ is raised by several degrees on annealing (figure \ref{fig:phasediagram}). Rather surprisingly, \it T\rm $\rm _c$ is also enhanced by annealing - more closely resembling previously reported values for polycrystalline samples~\cite{codopedsr} - accompanied by a noticeable broadening of the superconducting transition (figure \ref{fig:magnetisation}). Microprobe analysis reveals a loss of Sr on annealing, with annealed samples characterised by Sr-deficiency of up to 10\%. A similar effect is seen for samples grown out of Sn flux (figure \ref{fig:magnetisation}), with an enhancement of \it T\rm $\rm _c$ accompanied by a broadened superconducting transition. Sn impurities are known to substitute for Sr(Ba) in Sr(Ba)122 [ref.~\onlinecite{tinimpurities}], with a similar effect to the loss of Sr (the `spacer' inbetween the superconducting FeAs planes) on annealing. Possible consequences of this inter-plane depletion of Sr ions include effective electron-doping (similar to K-substitution~\cite{kdopedba}), a disturbance of the antiferromagnetic nesting, effective uniaxial strain~\cite{pressuresr2}, or a reduction in the Coulomb repulsion~\cite{itinerantelectrons}; all of which could be reasons for the enhancement in superconducting temperature.

Figure \ref{fig:phasediagram} shows a phase diagram for Co-doped Sr122 constructed from measurements of resistance, heat capacity, and magnetisation. The magnetic/structural transition temperature is obtained from the sharp feature in resistance (figure \ref{fig:resistivity}) $-$ traced up to the doping where the superconducting dome onsets, the sharp drop in magnetisation and the peak in heat capacity (figure \ref{fig:heatcapacity}) $-$ both of which are traced to lower dopings $x$=0.045. The superconducting transition temperature is obtained from the drop to zero resistance (figure \ref{fig:resistivity}) and the occurrence of superconducting screening in magnetisation (figure \ref{fig:magnetisation}). While the phase diagram of Co-doped Sr122 is superficially similar to Co-doped Ba122, vital differences are introduced due to the chemical tuning from Ba to Sr:

(i) the optimal \it T\rm $\rm _c$ in Co-doped Sr122 is only $\approx$70\% of the optimal \it T\rm $\rm _c$ common to hole-doped Ba122 [ref.~\onlinecite{canfield}] (fig.~\ref{fig:phasediagram}) while \it T\rm $\rm _N$ is over 40\% higher.

(ii) the region of magnetism in Co-doped Sr122 is further separated from the region of superconductivity, with less overlap - occurring only between 7.5\% and 8.5\% in Co-doped Sr122 as compared to the region between 2.5\% and 6\% in Co-doped Ba122.

(iii) on careful inspection, there is no resolvable splitting between the magnetic and structural transitions in Co-doped Sr122. Figure~\ref{fig:heatcapacity} shows the \it C\rm $\rm _p$ transition at various Co-dopings in Sr122. A comparison is made at a representative doping of Ba122 (chosen for a similar $\%$ suppression of $T\rm _N$), which reveals a distinct splitting in the transition. An absence of splitting in Co-doped Sr122 due to a smearing of the two transitions is ruled out by the greatly reduced width of the single transition in Co-doped Sr122.

(iv) the nature of the magnetic transition in Co-doped Sr122 is compared with that in Co-doped Ba122 by a study of the entropy change associated with the transition. Figure ~\ref{fig:heatcapacity}b) shows the integrated value of \it C$\rm _{mag}$\it /T \rm (red lines) as a function of temperature for Co-doped Sr122 and Co-doped Ba122, where \it C$\rm _{mag}$\it /T\rm = ($C\rm _p$ - \it C$\rm _{nonmag}$\it )/T\rm. Here \it C$\rm _{nonmag}$\it /T \rm is obtained from the measured $C\rm _p$ in a higher doped sample without a magnetic transition. The blue peak is the local value of \it C$\rm _{mag}$\it /T\rm. Whereas the entropy change at \it T\rm $\rm _N$ in Co-doped Sr122 shows a significantly more abrupt near-vertical change, consistent with the latent heat involved at a largely first order transition - the entropy change at \it T\rm $\rm _N$ in Co-doped Ba122 shows a more gradual change with an extended shoulder characteristic of a quasicontinuous transition.

All these differences between the Co-doped Sr122 and Ba122 phase diagrams can be associated with the split / coincident nature of the magnetic and structural transitions. First we turn to the origin of the difference in splitting between Co-doped Sr122 and Ba122. The quasicontinuous magnetic transition in Co-doped Ba122 occurs several Kelvin lower than the structural transition. In contrast, enhanced interlayer coupling in the more three-dimensional Sr122 leads to a higher magnetic transition that coincides with the structural transition to yield a single transition. Models that attempt to explain the splitting between magnetic and structural transitions in Ba122 and the 1111 family~\cite{nematicordertheory,orbital} have also discussed the effect of increased interlayer exchange as reducing the splitting between magnetic and structural transitions. While the application of external pressure was suggested as a way to tune the splitting, here we show that chemical tuning, by replacing Ba by Sr to increase the dimensionality of the 122 material in question, is an effective way of achieving this tuning. Next we turn to the difference in the order of the magnetic/structural transition in Co-doped Sr122 and Ba122. The first-order nature of the simultaneous magnetic/structural transition we observe in Co-doped Sr122 can be explained from symmetry as a consequence of the simultaneous occurrence of an Ising structural transition and an XY antiferromagnetic transition~\cite{symmetrytransition}. The first-order character of the transition in underdoped 122 materials where the magnetic and structural transitions coincide~\cite{krellner} has been characterised in detail, and the precise nature of the transition revealed by TEM measurements~\cite{jamesloudon}. Further, in contrast to the onset of the in-plane anisotropy in resistance in detwinned Co-doped Ba122 above \it T\rm $\rm _s$ and \it T\rm $\rm _N$ [ref.~\onlinecite{straindetwinning}] indicating fluctuations of the order parameter, these fluctuations are truncated in Co-doped Ca122, which exhibits a strongly first order transition in the undoped material~\cite{3d2dcafe2as2}. 

We propose that enhanced interactions at the split quasi-continuous transition in the lower dimensional Ba122 lead to enhanced superconducting temperatures, compared to the single first-order transition in higher dimensional Sr122, where $T_{\rm c}$ is suppressed, and the magnetic and superconducting domes further separated. Our finding in the 122 family of materials is of immediate relevance to the broader class of pnictide superconductors, such as the 1111 family, where reduced dimensionality and increased splitting between magnetic and structural transitions are accompanied by higher superconducting transition temperatures.

Work is supported by the EPSRC, Trinity College, the Royal Society and the Commonwealth Trust. The authors would like to thank D. Astill, J. Cooper, and M. Grosche for experimental assistance and C. Zentile, A. Bernevig, J. Loram, C. Geibel, and G. Lonzarich for useful discussions.

\bibliography{CoincidentTransition}

\end{document}